\documentclass[nofootinbib,onecolumn,showpacs,preprintnumbers,amsmath,amssymb]{revtex4}
\def\bea{\begin{eqnarray}}
\def\eea{\end{eqnarray}}
\newcommand{\be}{\begin{equation}}
\newcommand{\ee}{\end{equation}}

\newcommand{\bn}{\begin{eqnarray}}
\newcommand{\en}{\end{eqnarray}}

\newcommand{\p}{\partial}

\newcommand{\nn}{\nonumber}

\newcommand{\no}{\noindent}

\newcommand{\s}{\,\,\,\,}
\def\bea{\begin{eqnarray}}
\def\eea{\end{eqnarray}}

\usepackage{amsmath}
\usepackage{amssymb}
\usepackage{graphicx}
\usepackage{bm}
\usepackage{epstopdf}
\usepackage{slashed}

\usepackage[unicode=true,
 bookmarks=true,bookmarksnumbered=false,bookmarksopen=false,
 breaklinks=false,pdfborder={0 0 1},backref=false,colorlinks=false]
 {hyperref}

\begin{document}

\title {Unimodular gravity theory with external sources in a Lorentz-symmetry breaking scenario}

\author{L. H. C. Borges}
\email{luizhenriqueunifei@yahoo.com.br}

\author{D. Dalmazi}
\email{dalmazi@feg.unesp.br}

\affiliation{UNESP - Campus de Guaratinguet\'a - DFQ, Avenida Dr. Ariberto Pereira da Cunha 333, CEP 12516-410,
Guaratinguet\'a, SP, Brazil}

\begin{abstract}
This paper is dedicated to the study of interactions between stationary field sources for the linearized
unimodular gravity or WTDIFF theory in a model which exhibits Lorentz symmetry breaking due to the presence of
the linearized topological Chern-Simons term in $3 + 1$ dimensions, where the Lorentz symmetry breaking is
caused by a single background vector $v^{\mu}$. Since the background vector is very tiny, we treat it
perturbatively up to second order and we focus on physical phenomena which have no counterpart in standard
WTDIFF theory. We consider effects related to field sources describing point-like particles and cosmic strings.
We show that in a Lorentz violating
scenario the interaction between external sources lead to numerically different results
for linerarized Eintein-Hilbert (LEH) and WTDIFF theories,
however both results are qualitatively similar and can be equalized after a rescaling of 
the Lorentz breaking source term which makes an experimental distinction impossible at 
leading order in pertubation theory as far as 
point particles and cosmic strings are concerned.
\end{abstract}

\maketitle

\section{Introduction}
\label{I}

Theories with Lorentz-symmetry breaking have been under intensive investigation as an  attempt to find a
consistent description of quantum gravity. Most of these investigations have been done in the context of the
Standard Model extension (SME) \cite{LV1,LV2} that incorporates in the Standard Model the full set of
gauge-invariant, renormalizable Lorentz violation interactions. Some aspects of Lorentz violation have been
investigated, for example, in Maxwell electrodynamics \cite{LV3,LV4,LV5,LV6,LV7,LV8,LV9}, QED \cite{LV10,LV11},
linearized gravity \cite{G1,G2,G3,G4,G5,G6,G7,G8,G9,G10,G110,G120}, short-range experiments in pure gravity
\cite{G11,G12,G13,G14,G15,G16}.

Here we are interested in the issue of Lorentz violation in linearized gravity. There are two covariant descriptions of massless spin-2 particles in $3 + 1$ dimensions via a symmetric rank-2
tensor: the linearized Einstein-Hilbert (LEH) theory \cite{Hinter} and the Weyl plus transverse diffeomorphism
(WTDIFF) invariant model \cite{WT1,WT2,WT3,WT4}. The WTDIFF model is the linearized truncation of unimodular
gravity \cite{UG1,UG2,UG3,UG4}  which, in its turn, corresponds to the Einstein–Hilbert theory with the
replacement $g_{\mu\nu}\rightarrow {\hat{g}}_{\mu\nu}/(-{\hat{g}})^{1/4}$. The WTDIFF model can be obtained from
the usual LEH theory by the singular replacement $h_{\mu\nu}\rightarrow h_{\mu\nu}-\eta_{\mu\nu}h/4$, and the
external source is just the traceless piece of the energy-momentum tensor  $T_{\mu\nu}$, i.e.,
$T_{\mu\nu}\rightarrow T_{\mu\nu}-{\eta_{\mu\nu}}T/{4}$ \cite{abgv,EAMH} .

One remarkable feature of the WTDIFF theory  is the fact that this theory leads to the same results obtained in LEH theory for the interactions between external sources \cite{abgv,blas,EAMH}, at least, as far as the Lorentz symmetry of the linearized theory is preserved.
From this point of view it is natural to ask  whether such equivalence remains in a Lorentz violating scenario.
One might try to distinguish WTDIFF from LEH based on Lorentz violating phenomena as we pursue here. 
Moreover there are physical phenomena produced by the presence of external field sources,  with no counterpart 
in standard (Lorentz preserving) WTDIFF and LEH theories. Studies of this kind have not yet been considered in the
literature, to the best of the authors' knowledge and deserve investigations not only for their theoretical
aspects, but also because of their possible experimental relevance in the search for Lorentz symmetry breaking.

This paper is devoted to this subject, where we search for effects produced by external sources. At leading order in perturbation theory, we show that is not possible to distinguish the WTDIFF-LV and LEH-LV
theories through interactions between external sources. Specifically, we consider the WTDIFF Lagrangian modified by
the CPT breaking linearized topological Chern-Simons term in $3 + 1$ dimensions \cite{G1}, where the Lorentz symmetry
breaking is due to the presence of a single background vector $v^{\mu}$. Since the background vector is
very tiny, we treat it perturbatively up to leading order. We show that a spontaneous torque on a
classical rigid halter emerges and we investigate some phenomena due to the presence of cosmic strings and show that the
string can interact with a point-like particle as well as with another cosmic string in the Lorentz symmetry
breaking scenario considered.

The paper is structured as follows; in  Sect. \ref{II} we describe some general aspects of the models we deal with along
the paper and present our basic formula for the computation of the energy of the system. In Sect. \ref{III} we
consider effects due to the presence of point-like stationary particles.  Sect. \ref{IV} is dedicated to the
study of physical phenomena due to the presence of cosmic strings, and   Sect. \ref{conclusoes} is
dedicated to our final remarks and conclusions.

\section{Propagators and vacuum energy  in the presence of sources}
\label{II}

First, let us consider the following Lagrangian density in 3 + 1 dimensions
\begin{eqnarray}
\label{model} {\cal{L}}=-\frac{\sqrt{-g}}{2k^{2}}R-\frac{1}{2}\epsilon^{\mu\nu
k\lambda}v_{\mu}\Gamma^{\rho}_{\lambda\sigma} \left(\partial_{\nu}\Gamma^{\sigma}_{\rho
k}+\frac{2}{3}\Gamma^{\sigma}_{\nu\alpha}\Gamma^{\alpha}_{k\rho}\right) \ ,
\end{eqnarray}
where the first term is the well-known Einstein-Hilbert Lagrangian and the second one is the so called
topological Chern-Simons term in four dimensions, $g=\det g_{\mu\nu}$, $g_{\mu\nu}$ is the space-time metric, as
standard in the literature, we deffine the gravitational coupling as $k^{2}=8\pi G$, $R$ stands for the
scalar curvature,
\begin{eqnarray}
\label{cscalar}
R=g^{\mu\nu}\left(\partial_{\nu}\Gamma^{\lambda}_{\mu\lambda}-\partial_{\lambda}\Gamma^{\lambda}_{\mu\nu}+\Gamma^{\tau}_{\mu\lambda}
\Gamma^{\lambda}_{\tau\nu}-\Gamma^{\tau}_{\mu\nu}\Gamma^{\lambda}_{\tau\lambda}\right) \ ,
\end{eqnarray}
where
\begin{eqnarray}
\label{aconection}
\Gamma^{\lambda}_{\mu\nu}=\frac{1}{2}g^{k\lambda}\left(\partial_{\mu}g_{k\nu}+\partial_{\nu}g_{\mu
k}-\partial_{k}g_{\mu\nu}\right)\ ,
\end{eqnarray}
is the affine connection, $\epsilon^{\mu\nu k\lambda}$ stands for the Levi-Civita tensor
$\left(\epsilon^{0123}=1\right)$, and $v^{\mu}=\left(v^{0}, {\bf{v}}\right)$ is the background vector.

We will restrict our attention to a graviton propagating on a flat Minkowski geometry whose metric tensor is
$\eta^{\mu\nu}=(+,-.-,-)$. We use
\begin{eqnarray}
\label{weakfield} g_{\mu\nu}\left(x\right)=\eta_{\mu\nu}+kh_{\mu\nu}\left(x\right) \ ,
\end{eqnarray}
where $kh_{\mu\nu}$ is the symmetric spin-2 field and represents, as usually, a small pertubation around flat
Minkowski space-time.

The model (\ref{model}) at linearized level (quadratic in $h_{\mu\nu}$) is invariant under linearized general coordinate transformations,
\begin{eqnarray}
\label{gauget} \delta
h_{\mu\nu}\left(x\right)=\partial_{\mu}\xi_{\nu}\left(x\right)+\partial_{\nu}\xi_{\mu}\left(x\right) \ ,
\label{sym1}
\end{eqnarray}
where $\xi_{\nu}\left(x\right)$ are the gauge parameters. So, in order to avoid singularities it is necessary to fix
this gauge invariance, a common choice in the perturbative gravity literature is the De Donder gauge-fixing
term,
\begin{eqnarray}
\label{gauget1} {\cal{L}}_{gf}=\frac{1}{2\alpha}F_{\mu}F^{\mu}\ ,
\end{eqnarray}
with
\begin{eqnarray}
\label{gauget2} F^{\mu}=\partial_{\nu}\left(h^{\mu\nu}-\frac{1}{2}\eta^{\mu\nu}h\right) \ ,
\end{eqnarray}
and $h=\eta^{\mu\nu}h_{\mu\nu}$. Therefore, by using the weak field approximation (\ref{weakfield}) and taking
into account (\ref{gauget1}), the quadratic Lagrangian in the spin-2 field, reads
\begin{eqnarray}
\label{model2} {\cal{L}}_{EH-LV} &=&-\frac{1}{2}\left(\frac{1}{2}h_{\mu\nu}\Box h^{\mu\nu}-\frac{1}{2}h\Box
h+h\partial_{\mu}\partial_{\nu}h^{\mu\nu}
-h^{\mu\nu}\partial_{\mu}\partial^{\lambda}h_{\lambda\nu}\right)\nonumber\\
&
&+\frac{1}{2\alpha}\left(-h^{\mu\nu}\partial_{\mu}\partial^{\lambda}h_{\lambda\nu}+h^{\mu\nu}\partial_{\mu}\partial_{\nu}h
-\frac{1}{4}h\Box h\right)\nonumber\\
& &-\frac{k^{2}}{4}\left[\epsilon^{\mu\nu k\lambda}v_{\mu}\left(h_{\lambda}^{\ \rho}\Box\partial_{\nu}h_{\rho
k}-
h_{\lambda\rho}\partial_{\nu}\partial^{\rho}\partial^{\sigma}h_{\sigma k}\right)\right]\nonumber\\
& &+\frac{1}{2}k\, h_{\mu\nu}T^{\mu\nu} \ ,
\end{eqnarray}
where in the last line we have inserted an external source, $T^{\mu\nu}$, that must be conserved
$\partial_{\mu}T^{\mu\nu}=0$ in order that the source term $\int d^4 x \, h_{\mu\nu}T^{\mu\nu}$ be invariant
under linearized reparametrizations (\ref{sym1}).

Now let us jump to the WTDIFF model. There are only two ways of describing massless spin-2 particles covariantly
in terms of a symmetric rank-2 tensor. Besides the usual linearized Einstein-Hilbert theory ${\cal L}_{EH}$
which is the far more popular theory, we have a model invariant under  Weyl (W) transformations and transverse
reparametrizations (TDIFF). There are elderly \cite{u,hu} and more recent \cite{abgv,blas}  references on this subject
. In practice the so called WTDIFF model can be obtained from the Einstein-Hilbert theory
(\ref{model2}) in $D=3+1$  via the singular replacement by a traceless field, i.e.,  $h_{\mu\nu} \to h_{\mu\nu}-
\eta_{\mu\nu}h/4$. From (\ref{model2}) we obtain

\begin{eqnarray}
\label{model3} {\cal{L}}_{WT-LV} &=&-\frac{1}{2}\left(\frac{1}{2}h_{\mu\nu}\Box h^{\mu\nu}-\frac{3}{16}h\Box h +
\frac 12 h\partial_{\mu}\partial_{\nu}h^{\mu\nu}
-h^{\mu\nu}\partial_{\mu}\partial^{\lambda}h_{\lambda\nu}\right)\nonumber\\
&
&+\frac{1}{2\alpha}\left(-h^{\mu\nu}\partial_{\mu}\partial^{\lambda}h_{\lambda\nu}+h^{\mu\nu}\partial_{\mu}\partial_{\nu}h
-\frac{1}{4}h\Box h\right)\nonumber\\
& &-\frac{k^{2}}{4}\left[\epsilon^{\mu\nu k\lambda}v_{\mu}\left(h_{\lambda}^{\ \rho}\Box\partial_{\nu}h_{\rho
k}-
h_{\lambda\rho}\partial_{\nu}\partial^{\rho}\partial^{\sigma}h_{\sigma k}\right)\right]\nonumber\\
& &+\frac{1}{2}k\, h_{\mu\nu}\hat{T}^{\mu\nu} \quad .
\end{eqnarray}

\no Notice that the singular redefinition has  automatically generated a redefinition of the source:

\be T_{\mu\nu} \to \hat{T}_{\mu\nu} = T_{\mu\nu} - \eta_{\mu\nu} \frac T4 \quad . \label{that} \ee

\no The new source is defined out of the old conserved source $T_{\mu\nu}$  but it is not conserved  by itself.
It satisfies instead the constraints:

\be \eta^{\mu\nu}\hat{T}_{\mu\nu} =0 \quad ; \quad  \p^{\mu}\hat{T}_{\mu\nu} = -\p_{\nu}T/4 \quad .
\label{cthat}\ee

\no The new constraints make the source term $\int d^4 x \, h_{\mu\nu}{\hat
T}^{\mu\nu}$ invariant under  transverse reparametrizations and Weyl transformations:

\be \delta h_{\mu\nu} = \p_{\mu}\xi_{\nu}^T + \p_{\nu} \xi_{\mu}^T + \eta_{\mu\nu} \phi \quad ; \quad
\p^{\mu}\xi_{\mu}^T = 0 \quad ,  \label{sym2} \ee

\no which are the symmetries of ${\cal L}_{WT-LV}$, except for the gauge fixing term of course where we have not
made any  field redefinition, otherwise the Weyl symmetry would not be broken by such term. We have made use of
the fact that the Lorentz violating term is invariant under Weyl transformations and (arbitrary)
reparametrizations.

Now we are going to write down  the propagators corresponding to ${\cal L}_{EH-LV}$ and ${\cal L}_{WT-LV}$ and
the respective source-source term. The presence of the LV term makes the calculation complicate. Fortunately, it
has been worked out before in \cite{G3}. Both Lagrangians (\ref{model2}) and (\ref{model3}) fit in the form

\be {\cal L} = C_1 (\p^{\mu}h_{\mu\nu})^2 + C_2 \, \p^{\mu}h \p^{\alpha}h_{\alpha\mu} + h\, C_3(\Box) h +
h^{\mu\nu}C_4(\Box) h_{\mu\nu} + h_{\lambda\alpha} \Box \theta^{\rho\alpha}S^{\nu\lambda} h_{\rho\nu} \quad ,
\label{ele} \ee

\no where $C_1, C_2 $ are constants, $C_3(\Box), C_4(\Box)$ are functions of $\Box$ and
$S^{\nu\lambda}=k^{2}\epsilon^{\nu\lambda\beta\sigma}v_{\beta}\p_{\sigma}$. If we write down (\ref{ele}) in the form ${\cal L} = h^{\mu\nu}G_{\mu\nu,\alpha\beta}h^{\alpha\beta}$, the differential operator $G_{\mu\nu,\alpha\beta}$ can be decomposed in terms of a basis of operators \cite{G3} displayed in our appendix. Suppressing
indices we have 

\bea G && = C_4 P_{SS}^{(2)} + \left(C_4- \frac{C_1 \Box}2 \right) P_{SS}^{(1)} + \left\lbrack C_3 + C_4 - (C_1 +
C_2)\Box \right\rbrack P_{WW}^{(0)} +  (3 \, C_3 + C_4) P_{SS}^{(0)}  \nn \\ &&  + \sqrt{3}\left( C_3 -
\frac{C_2\Box}2 \right)(P_{SW}^{(0)}+  P_{WS}^{(0)}) - \frac {\Box}2 (S_{\sigma}+S_{\tau}) \ . \label{Gc}   \eea

\no Using (\ref{Gmenos}) we have the inverse operator

\bea G^{-1}&& = \frac{P_{SS}^{(2)}}{C_4} + \frac{P_{SS}^{(1)}}{C_4- \frac{C_1 \Box}2}  + \frac{3 \, C_3 +
C_4}{K^{(0)}} P_{WW}^{(0)}  + \frac{\left\lbrack C_3 + C_4 - (C_1+C_2)\Box \right\rbrack }{K^{(0)}}P_{SS}^{(0)}
\nn \\ && - \sqrt{3}\left( C_3 - \frac{C_2\Box}2 \right)(P_{SW}^{(0)}+  P_{WS}^{(0)}) + G^{-1}_{LV} \ ,
\label{gmenosc} \eea

\no where the spin-0 determinant $K^{(0)}$ and the  Lorentz-violation piece are given respectively by

\be K^{(0)} = \left\lbrack C_3 + C_4 - (C_1+C_2)\Box \right\rbrack \left( 3 \, C_3 + C_4 \right) - 3\left( C_3 -
\frac{C_2\Box}2 \right)^2 \quad , \label{kzero} \ee

\be G^{-1}_{LV} = \frac{4\left(iS_{\tau}+2S_{\tau}^{2}\right)}{\Box\left[1+4f\left(p\right)\right]}+
\frac{4\left(iS_{\sigma}+2S_{\sigma}^{2}\right)}{\Box\left[1+16f\left(p\right)\right]} \quad , \label{glv} \ee

\no where $f(p)$ is given in (\ref{efe}) and the operators $S_{\sigma}$, $S_{\sigma}^2$, $S_{\tau}$ and $S_{\tau}^2$ are also displayed in the appendix.  In the cases of EH and WTDIFF we have respectively

\bea C_1 &&= 1 + \frac 1{2\alpha} \quad ; \quad C_2= -(1 + \frac 1{2\alpha}) \quad ; \quad C_3 = -\frac{\Box}2 -
\frac{\Box}{8\alpha} \quad ; \quad  C_4 = \frac{\Box}2 \quad  ; \quad K^{(0)} = \frac{\Box^2}{8\alpha} \, , \label{ceh} \\
C_1 &&= 1 + \frac 1{2\alpha} \quad ; \quad C_2= -\frac 12 - \frac 1{2\alpha} \quad ; \quad C_3 = -\frac{3\Box}{16}
- \frac{\Box}{8\alpha} \quad ; \quad  C_4 = \frac{\Box}2 \quad  ; \quad K^{(0)} = \frac{\Box^2}{32\alpha} \ .
\label{cwt} \eea

\no Explicitly we have:

\bea G^{-1}_{EH-LV} &=&  \frac{P_{SS}^{(2)}}{\Box/2} - \frac{P_{SS}^{(1)}}{\Box/(4\alpha)} - \left(\frac 3{\Box} +
\frac{8\alpha}{\Box}\right)P_{WW}^{(0)} - \frac{P_{SS}^{(0)}}{\Box} - \frac{\theta\omega +\omega\theta}{\Box} +
G_{LV}^{-1} \, , \label{gmenoseh} \\ G^{-1}_{WT-LV} &=&  \frac{P_{SS}^{(2)}}{\Box/2} -
\frac{P_{SS}^{(1)}}{\Box/(4\alpha)} - \left(\frac{12}{\Box} + \frac{2\alpha}{\Box}\right)P_{WW}^{(0)} -
\left(\frac 4{\Box} + \frac{6\alpha}{\Box} \right)P_{SS}^{(0)} - \left(\frac 4{\Box} +
\frac{2\alpha}{\Box}\right)(\theta\omega +\omega\theta ) + G_{LV}^{-1} \, . \label{gmenoswt} \eea

\no Notice that, since $G^{-1}_{LV}$ only depends upon the Lorentz violating term and the coefficient $C_4$ in the operator $G$, see (\ref{G}) and (\ref{Gmenos}), we end up with the same Lorentz violating term in $G^{-1}$ for both EH and
WTDIFF models. After we saturate $G^{-1}_{EH-LV}$ with conserved sources it turns out that all terms containing the derivatives $\omega_{\mu\nu}$ drop out. We have in momentum space:

\bea T^{\mu\nu}(-p)\left(G^{-1}_{EH-LV}\right)_{\mu\nu,\alpha\beta}T^{\alpha\beta}(p) && =  T^{\mu\nu}(-p)\left\lbrack \left( \frac {2\,  P_{SS}^{(2)}}{p^2} - \frac{P_{SS}^{(0)}}{p^2}\right) +  G_{LV}^{-1} \right\rbrack_{\mu\nu,\alpha\beta}T^{\alpha\beta}(p)\nn\\ &&= T^{\mu\nu}(-p)\left\lbrack \frac{\eta_{\mu\alpha}\eta_{\nu\beta} + \eta_{\mu\beta}\eta_{\nu\alpha}}{p^2} - \left(\frac 23 + \frac 13 \right) \frac{\eta_{\mu\nu}\eta_{\alpha\beta}}{p^2} + G_{LV}^{-1}  \right\rbrack_{\mu\nu,\alpha\beta}T^{\alpha\beta}(p) \, , \label{tgteh1} \\
&&=  T^{\mu\nu}(-p)\left\lbrack \frac{\eta_{\mu\alpha}\eta_{\nu\beta} + \eta_{\mu\beta}\eta_{\nu\alpha}}{p^2} - \frac{\eta_{\mu\nu}\eta_{\alpha\beta}}{p^2} \right\rbrack_{\mu\nu,\alpha\beta}T^{\alpha\beta}(p)\nonumber\\
&
& +  T^{\mu\nu}(-p) \left( G_{LV}^{-1} \right)_{\mu\nu,\alpha\beta}T^{\alpha\beta}(p)  \label{tgteh2} \ .  \eea 

\no Notice the factor $1/3$ in (\ref{tgteh1}) which comes from $P_{SS}^{(0)}$, it is  responsible for the vDVZ \cite{vdv,zak} mass discontinuity, since it does not appear in the  propagator of the massive theory, no matter how small is the mass. 

Now, if we saturate  $G^{-1}_{WT-LV}$ with the traceless sources $\hat{T}_{\mu\nu}$, which satisfy  (\ref{cthat}), we have once again no contribution from the spin-1 sector, though this is more subtle now:

\be  \hat{T}^{\mu\nu}(-p)\left( P_{SS}^{(1)}\right)_{\mu\nu,\alpha\beta}\hat{T}^{\alpha\beta}(p)= 2\, \hat{T}^{\mu\nu}(-p) \theta_{\mu\alpha}\omega_{\nu\beta}\hat{T}^{\alpha\beta}(p)= \frac 12 \hat{T}^{\mu\nu}(-p) \theta_{\mu\alpha}p_{\nu}p^{\alpha}T = 0 \, , \label{pss1} \ee

\no where the last equality holds due to the transverse property of $\theta_{\mu\alpha}$. It turns out that adding up all spin-0 contributions we derive the known result \cite{UG2} that no one can tell a difference between the source-source terms in LEH and WTDIFF theories as far as Lorentz preserving terms are considered, namely from (\ref{gmenoswt}),

\bea  \hat{T}^{\mu\nu}(-p)\left(G^{-1}_{WT-LV}\right)_{\mu\nu,\alpha\beta} \hat{T}^{\alpha\beta}(p) && =   \hat{T}^{\mu\nu}(-p)\left\lbrack  \frac {2\,  P_{SS}^{(2)}}{p^2} - \left( \frac{12}{\, p^2} + \frac{4}{3 \, p^2} - \frac 8{p^2}\right)\omega\omega +  G_{LV}^{-1} \right\rbrack_{\mu\nu,\alpha\beta} \hat{T}^{\alpha\beta}(p) \nn\\ && =  T^{\mu\nu}(-p)\left\lbrack \frac{\eta_{\mu\alpha}\eta_{\nu\beta} + \eta_{\mu\beta}\eta_{\nu\alpha}}{p^2} - \frac{\eta_{\mu\nu}\eta_{\alpha\beta}}{p^2} \right\rbrack_{\mu\nu,\alpha\beta}T^{\alpha\beta}(p)\nonumber\\ 
&
&+\hat{T}^{\mu\nu}(-p) \left( G_{LV}^{-1} \right)_{\mu\nu,\alpha\beta}\hat{T}^{\alpha\beta}(p) \quad .  \label{tgtwt}\eea

\no Notice that the only difference between (\ref{tgteh2}) and (\ref{tgtwt}) is the replacement of $T^{\mu\nu}$ by $\hat{T}^{\mu\nu}$  in the Lorentz violating term. 

Paving the way for the next section we present here the formulae for the vacuum energy.
Since we have a quadratic Lagrangian in the field variables, the contribution due to the stationary sources to the vacuum energy of the system is given in the WTDIFF case by \cite{Zee}
\begin{eqnarray}
\label{energy1WT1} E_{WT-LV}=\frac{k^{2}}{4t}\int\int d^{4}x \ d^{4}y \
{\hat{T}}^{\alpha\beta}\left(x\right){\hat{D}}_{\alpha\beta,\gamma\delta}\left(x,y\right){\hat{T}}^{\gamma\delta}\left(y\right)
\ ,
\end{eqnarray}
where $t$ is a time variable and the limit $t\rightarrow\infty$ is implicit. The integration in $y^{0}$ is from
$-t/2$ to $t/2$.  We have introduced the propagator in the coordinate space
\begin{eqnarray}
\label{propagatorWT}
{\hat{D}}_{\alpha\beta,\gamma\delta}\left(x,y\right)&=&\int\frac{d^{4}p}{\left(2\pi\right)^{4}}{\hat{D}}_{\alpha\beta,\gamma\delta}\left(p\right)e^{-ip\cdot\left(x-y\right)}
\ .
\end{eqnarray}
Splitting the momentum space propagator into Lorentz preserving and Lorentz violating parts we have from (\ref{tgtwt}):
\begin{eqnarray}
\label{energy1WT3} E_{WT-LV}=\frac{k^{2}}{4t}\int\frac{d^{4}p}{\left(2\pi\right)^{4}}\int\int d^{4}x \ d^{4}y \
e^{-ip\cdot\left(x-y\right)}\Bigl[{{T}}^{\alpha\beta}\left(x\right){{D}}_{\alpha\beta,\gamma\delta}^{(0)}\left(p\right){{T}}^{\gamma\delta}\left(y\right)+{\hat{T}}^{\alpha\beta}\left(x\right){{D}}_{\alpha\beta,\gamma\delta}^{(LV)}\left(p\right){\hat{T}}^{\gamma\delta}\left(y\right)\Bigr]
\ ,
\end{eqnarray}

\no where 

\be {{D}}_{\alpha\beta,\gamma\delta}^{(0)}\left(p\right)   =  \frac{\eta_{\alpha\gamma}\eta_{\beta\delta} + \eta_{\alpha\delta}\eta_{\beta\gamma}}{p^2} - \frac{\eta_{\alpha\beta}\eta_{\gamma\delta}}{p^2} \,\, ; \,\, {{D}}_{\alpha\beta,\gamma\delta}^{(LV)}\left(p\right) =  \frac{4\left(iS_{\tau}+2S_{\tau}^{2}\right)_{\alpha\beta,\gamma\delta}}{p^2\left[1+4f\left(p\right)\right]}+
\frac{4\left(iS_{\sigma}+2S_{\sigma}^{2}\right)_{\alpha\beta,\gamma\delta}}{p^2\left[1+16f\left(p\right)\right]}  \, . \label{split} \ee

\no From (\ref{energy1WT3}) we can compute the interaction energy between different spin-2 field sources
for the WTDIFF-LV model. The  same expression (\ref{energy1WT3}) can be used for the EH-LV case if we replace  $\hat{T}^{\mu\nu} \to T^{\mu\nu}$ in the Lorentz violating term.

As the Lorentz symmetry breaking must be very tiny, the background vector must be small. Therefore, along the
paper we will treat $v^{\mu}$ perturbatively up to second order, which is the leading order in the background
vector.

\section{Point-like particles}
\label{III}
In this section we study the interaction energy between two stationary point-like particles in 3 + 1 dimensions.
The external source which describes this system is given by
\begin{eqnarray}
\label{corre1} \left[{{T}}^{\alpha\beta}({x})\right]^{I}=M_{1}\eta^{\alpha 0}\eta^{\beta 0}\delta^{3}\left({\bf
x}-{\bf a}_ {1}\right)+ M_{2}\eta^{\alpha 0}\eta^{\beta 0}\delta^{3}\left({\bf x}-{\bf a}_ {2}\right) \ ,
\end{eqnarray}
\begin{eqnarray}
\label{corre11} \left[{\hat{T}}^{\alpha\beta}({x})\right]^{I}=M_{1}\left(\eta^{\alpha 0}\eta^{\beta
0}-\frac{\eta^{\alpha\beta}}{4}\right)\delta^{3}\left({\bf x}-{\bf a}_ {1}\right)+ M_{2}\left(\eta^{\alpha
0}\eta^{\beta 0}-\frac{\eta^{\alpha\beta}}{4}\right)\delta^{3}\left({\bf x}-{\bf a}_ {2}\right) \ ,
\end{eqnarray}
where the location of the particles are specified by the vectors ${\bf a}_{1}$ and ${\bf a}_{2}$, and the parameters $M_{1}$ and $M_{2}$ are the particles masses.

Substituting Eqs. (\ref{corre1}) and (\ref{corre11}) into Eq. (\ref{energy1WT3}) and discarding the
self-interacting contributions (that is, the interactions of a given point particle with itself), we obtain
\begin{eqnarray}
\label{energy2}
E_{WT-LV}^{I}&=&\frac{k^{2}M_{1}M_{2}}{2t}\int\frac{d^{4}p}{\left(2\pi\right)^{4}}\int\int d^{4}x \ d^{4}y \ \delta^{3}\left({\bf x}-{\bf a}_ {1}\right)\delta^{3}\left({\bf y}-{\bf a}_ {2}\right)e^{-ip\cdot\left(x-y\right)}\Bigl[D_{00,00}^{(0)}\left(p\right)+\frac{9}{16}D_{00,00}^{(LV)}\left(p\right)\nonumber\\
&
&+\frac{3}{16}\Bigl(D_{00,11}^{(LV)}\left(p\right)+ D_{00,22}^{(LV)}\left(p\right)+D_{00,33}^{(LV)}\left(p\right)\Bigr)+\frac{3}{16}\Bigl(D_{11,00}^{(LV)}\left(p\right)+ D_{22,00}^{(LV)}\left(p\right)+D_{33,00}^{(LV)}\left(p\right)\Bigr)\nonumber\\
&
&+\frac{1}{16}\Bigl(D_{11,11}^{(LV)}\left(p\right)+ D_{22,22}^{(LV)}\left(p\right)+D_{33,33}^{(LV)}\left(p\right)+D_{11,22}^{(LV)}\left(p\right)+ D_{22,11}^{(LV)}\left(p\right)+D_{11,33}^{(LV)}\left(p\right)\nonumber\\
& &+D_{33,11}^{(LV)}\left(p\right)+ D_{22,33}^{(LV)}\left(p\right)+D_{33,22}^{(LV)}\left(p\right)\Bigr)\Bigr] \
.
\end{eqnarray}

By using the Eqs. (\ref{split}), (\ref{OPERATOR1}), (\ref{OPERATOR2}), (\ref{OPERATOR3}), (\ref{OPERATOR4}), (\ref{OPERATOR5}), (\ref{OPERATOR6}), (\ref{OPERATOR7}) up to second order in $v^{\mu}$, computing the integrals in the
following order: $d^{3}{\bf x}$, $d^{3}{\bf y}$, $dx^{0}$, $dp^{0}$ and $dy^{0}$, using the Fourier
representation for the Dirac delta function $\delta(p^{0})=\int dx^{0}/(2\pi)\exp(-ipx^{0})$, and identifying
the time interval as $t=\int_{-\frac{t}{2}}^{\frac{t}{2}}dy^{0}$, we arrive at
\begin{eqnarray}
\label{energy3} E_{WT-LV}^{I}&=&-\frac{k^{2}M_{1}M_{2}}{2}\Bigl[\int\frac{d^{3}{\bf p}}{(2\pi)^{3}}\frac{e^{i{\bf
p}\cdot{\bf a}}}{{\bf p}^2} -10k^{4}\left({\bf{v}}\cdot {\bf{\nabla}}_{{\bf a}}\right)^{2}\int\frac{d^{3}{\bf
p}}{(2\pi)^{3}}
\frac{e^{i{\bf p}\cdot{\bf a}}}{{\bf p}^2}\nonumber\\
& &+2k^{4}\left[(v^{0})^{2}-5{\bf{v}}^{2}\right]\int\frac{d^{3}{\bf p}}{(2\pi)^{3}}e^{i{\bf p}\cdot{\bf
a}}\Bigr]\ ,
\end{eqnarray}
where ${\bf a}={\bf a}_{1}-{\bf a}_{2}$ is the distance between the two massive particles and we defined the
differential operator
\begin{eqnarray}
\label{deffo} {\bf{\nabla}}_{{\bf a}}=\left(\frac{\partial}{\partial {a^{1}}},\frac{\partial}{\partial {a^{2}}},
\frac{\partial}{\partial {a^{3}}}\right) \ .
\end{eqnarray}

The last term inside the brackets of Eq. (\ref{energy3}) is the Dirac delta function
$\delta^{3}\left({\bf{a}}\right)$ and provided that ${\bf{a}}$ is nonzero, this term vanishes.

Using the fact that
\begin{eqnarray}
\label{int1} \int\frac{d^{3}{\bf p}}{(2\pi)^{3}}\frac{e^{i{\bf p}\cdot{\bf a}}}{{\bf p}^{2}}=\frac{1}{4\pi a}\ ,
\end{eqnarray}
whrere $a=|{\bf a}|$, and performing some manipulations, we obtain
\begin{eqnarray}
\label{energy4} E_{WT-LV}^{I}=-\frac{GM_{1}M_{2}}{a}\left[1-10\left(\frac{8\pi
G}{a}\right)^{2}\left(\frac{3\left({\bf{v}}\cdot{\bf{a}}\right)^{2}} {a^{2}}-{\bf {v}}^{2}\right)\right] \ .
\end{eqnarray}

Equation (\ref{energy4}) is a perturbative result and gives the interaction energy between two point-like
particles mediated by the spin-2 field with the specific Lorentz violating coupling contained in Eq.
(\ref{model2}). The $v^{\mu}$ dependent contribution in (\ref{energy4}) is a correction to the usual
gravitational interaction due the Lorentz symmetry breaking, leading to an anisotropic interaction between the
particles. If we take $v^{\mu}\rightarrow 0$, the expression reduces to the standard gravitational interaction.
The same happens for the case where $v^{\mu}=\left(v^{0},0\right)$, i.e., if ${\bf{v}}=0$. However, for the
particular case where the distance vector ${\bf{a}}$ is perpendicular to the background vector ${\bf{v}}$,
Eq.(\ref{energy4}) still exhibits effects due to Lorentz symmetry breaking,
\begin{eqnarray}
\label{energy5} E_{WT-LV}^{I}\left({\bf v}\cdot{\bf a}=0\right)=-\frac{GM_{1}M_{2}}{a}\left[1+10\left(\frac{8\pi
G}{a}\right)^{2}{\bf {v}}^{2}\right] \ .
\end{eqnarray}

The force between the two particles can be calculated from Eq. (\ref{energy4}), resulting in
\begin{eqnarray}
\label{force1}
&{\bf{F}}_{WT-LV}^{I}&=-{\bf{\nabla}}_{{\bf a}}E_{WT-LV}^{I}\nonumber\\
& &=-\frac{GM_{1}M_{2}}{a^{2}}\Biggl\{\left[1+30\left(\frac{8\pi G}{a}\right)^{2}\left({\bf{v}}^{2}-
\frac{5\left({\bf{v}}\cdot{\bf{a}}\right)^{2}}{a^{2}}\right)\right]{\hat{a}}+60\left(\frac{8\pi G}{a}\right)^{2}
\frac{\left({\bf{v}}\cdot{\bf{a}}\right)}{a}{\bf{v}}\Biggr\} \ ,
\end{eqnarray}
where ${\hat{a}}$ is an unit vector pointing in the direction of the vector ${\bf{a}}$. The interaction force in
(\ref{force1}) shows us more explicitly the anisotropies generated by the Lorentz symmetry breaking. 

We remark that the authors of \cite{BaileyHavert} have considered the effect of Lorentz and CPT breaking terms of mass dimension five (three derivatives) on a couple of moving point masses. Their Lorentz violating terms include ours, see (\ref{model2}). In the static limit their inverse cubic force contribution $F \sim a^{-3}$ disappears in agreement with (\ref{force1}) where our 
leading Lorentz breaking contribution only appears at  order \footnote{The case of moving point particles as well as the use of more general Lorentz violating terms is now in progress.} $a^{-4}$.

An important consequence of the anisotropies in expression (\ref{energy4}) is the emergence of a spontaneous
torque on an rigid halter. In order to investigate this effect we consider a typical rigid hater composed by two
particles of masses $M_{1}$ and $M_{2}$ respectively, placed at the positions ${\bf a}_{1}={\bf R}+\frac{{\bf
A}}{2}$ and ${\bf a}_{2}={\bf R}-\frac{{\bf A}}{2}$, where we take the distance vector ${\bf A}$ fixed (and
small). From Eq.\,(\ref{energy4}), we obtain
\begin{eqnarray}
\label{energy6} E_{WT-LV}^{halter}=-\frac{GM_{1}M_{2}}{A}\left[1-10\left(\frac{8\pi G}{A}\right)^{2}{\bf{v}}^{2}
\left(3\cos^{2}(\theta)-1\right)\right] \ ,
\end{eqnarray}
where $A=\mid {\bf A}\mid$ and $\theta$ is the angle between the vectors ${\bf A}$ and ${\bf v}$. Notice that
$0\leq\theta\leq 2\pi$. The energy (\ref{energy6}) leads to a spontaneous torque on the rigid halter, as
follows:
\begin{eqnarray}
\label{torque1}
&\tau_{WT-LV}^{halter}&=-\frac{\partial E_{WT-LV}^{halter}}{\partial\theta}\nonumber \\
& &=30\frac{GM_{1}M_{2}}{A}\left(\frac{8\pi G}{A}\right)^{2}{\bf{v}}^{2}\sin\left(2\theta\right) \ .
\end{eqnarray}

This spontaneous torque on the rigid halter is an exclusive effect due to the Lorentz violating background. If,
$v^{\mu}=0$ the torque vanishes, as it should, as well as for the specific cases $\theta=0,\pi,\pi/2,2\pi$. When
$\theta=\pi/4$, the torque exhibits a maximum value. 

For the LEH-LV theory we proceed as previously, replacing
${\hat{T}}^{\alpha\beta}\left(x\right)\rightarrow{{T}}^{\alpha\beta}\left(x\right)$ and
${\hat{T}}^{\gamma\delta}\left(y\right)\rightarrow{{T}}^{\gamma\delta}\left(y\right)$ in the second term between
brackets on the right hand side of the Eq. (\ref{energy1WT3}), obtaining for the interaction energy between two
point-like particles up to second order in $v^{\mu}$, the result
\begin{eqnarray}
\label{energy4EH} 
E_{EH-LV}^{I}&=&\frac{k^{2}M_{1}M_{2}}{2t}\int\frac{d^{4}p}{\left(2\pi\right)^{4}}\int\int d^{4}x \ d^{4}y \ \delta^{3}\left({\bf x}-{\bf a}_ {1}\right)\delta^{3}\left({\bf y}-{\bf a}_ {2}\right)e^{-ip\cdot\left(x-y\right)}\Bigl[D_{00,00}^{(0)}\left(p\right)+D_{00,00}^{(LV)}\left(p\right)\Bigr]\nonumber\\
&=&-\frac{GM_{1}M_{2}}{a}\left[1-16\left(\frac{8\pi
G}{a}\right)^{2}\left(\frac{3\left({\bf{v}}\cdot{\bf{a}}\right)^{2}} {a^{2}}-{\bf {v}}^{2}\right)\right] \ .
\end{eqnarray}

Comparing (\ref{energy4}) with (\ref{energy4EH})  we verify  that for WTDIFF-LV and LEH-LV theories, the interaction energy between
two point particles is numerically different already at second order in the background vector $v^{\mu}$. An opposite situation occurs in standard WTDIFF and LEH theories where the energies are the same ones. The Lorentz violating
terms in (\ref{energy4}) and (\ref{energy4EH}) are qualitatively similar and we could turn the
overall factor 10 into 16 by scaling $v^{\mu}$ into 
$(4/\sqrt{10}) v^{\mu}$, at least at leading order in perturbation theory. It is not clear whether such simple rescaling will fit the remaining terms beyond the quadratic approximation in $v^{\mu}$. This is under investigation.

\section{Cosmic strings}
\label{IV}

Opposite to the point-like particles of last section we now focus on long size objects, namely, cosmic strings.
Their contribution to the interaction energy, at leading order in perturbation theory, can only appear due to the Lorentz violating terms.

We start this section considering the interaction energy between a point-like particle and an cosmic string,
both of them stationary. The cosmic string shall be taken to flow parallel to the $z$-axis, along the straight
line located at ${\bf A}=(A^{1},A^{2},0)$. The point-like particle is concentrated at position
${\bf{s}}$. This system is described by the external source,
\begin{eqnarray}
\label{corre2} \left[T^{\alpha\beta}({x})\right]^{II}=M\eta^{\alpha 0}\eta^{\beta 0}\delta^{3}\left({\bf x}-{\bf
s}\right)+ \mu\left(\eta^{\alpha 0}\eta^{\beta 0}-\eta^{\alpha 3}\eta^{\beta
3}\right)\delta^{2}\left({\bf{x}}_{\perp}-{\bf A}\right) \ ,
\end{eqnarray}
where the first term on the right hand side of the above equation stands for the external field source produced
by the point-like charge, and the second one is the source produced by the cosmic string \cite{CosmicS}. The
mass parameter $M$ and the linear mass density $\mu$ are the coupling constants between the field and the delta
functions, and ${\bf x}_{\perp}=(x^{1},x^{2},0)$, is the position vector perpendicular to the cosmic string. For
the traceless source we have
\begin{eqnarray}
\label{corre22} \left[{\hat{T}}^{\alpha\beta}({x})\right]^{II}=M\left(\eta^{\alpha 0}\eta^{\beta
0}-\frac{\eta^{\alpha\beta}}{4}\right)\delta^{3}\left({\bf x}-{\bf s}\right)+ \mu\left(\eta^{\alpha
0}\eta^{\beta 0}-\eta^{\alpha 3}\eta^{\beta
3}-\frac{\eta^{\alpha\beta}}{2}\right)\delta^{2}\left({\bf{x}}_{\perp}-{\bf A}\right)\ .
\end{eqnarray}

Substituting the sources (\ref{corre2}) and (\ref{corre22}) in (\ref{energy1WT3}) and discarding the
self-interacting terms, which do not contribute to the interaction force between the string and the particle
(the self-interacting terms are proportional to $M^{2}$ or $\mu^{2}$), we have
\begin{eqnarray}
\label{energy21}
E_{WT-LV}^{II}&=&\frac{k^{2}M\mu}{2t}\int\frac{d^{4}p}{\left(2\pi\right)^{4}}\int\int d^{4}x \ d^{4}y \ \delta^{3}\left({\bf x}-{\bf s}\right)\delta^{2}\left({\bf{y}}_{\perp}-{\bf A}\right)e^{-ip\cdot\left(x-y\right)}\Bigl[D_{00,00}^{(0)}\left(p\right)-D_{00,33}^{(0)}\left(p\right)+\frac{3}{8}D_{00,00}^{(LV)}\left(p\right)\nonumber\\
&
&+\frac{3}{8}\Bigl(D_{00,11}^{(LV)}\left(p\right)+ D_{00,22}^{(LV)}\left(p\right)+D_{00,33}^{(LV)}\left(p\right)\Bigr)+\frac{1}{8}\Bigl(D_{11,00}^{(LV)}\left(p\right)+ D_{22,00}^{(LV)}\left(p\right)+D_{33,00}^{(LV)}\left(p\right)\Bigr)\nonumber\\
&
&+\frac{1}{8}\Bigl(D_{11,11}^{(LV)}\left(p\right)+ D_{22,22}^{(LV)}\left(p\right)+D_{33,33}^{(LV)}\left(p\right)+D_{11,22}^{(LV)}\left(p\right)+ D_{22,11}^{(LV)}\left(p\right)+D_{11,33}^{(LV)}\left(p\right)\nonumber\\
& &+D_{33,11}^{(LV)}\left(p\right)+
D_{22,33}^{(LV)}\left(p\right)+D_{33,22}^{(LV)}\left(p\right)\Bigr)-\frac{3}{4}D_{00,33}^{(LV)}\left(p\right)-\frac{1}{4}\left(D_{11,33}^{(LV)}\left(p\right)+D_{22,33}^{(LV)}\left(p\right)+
D_{33,33}^{(LV)}\left(p\right)\right)\Bigr] \ ,
\end{eqnarray}
where the integration limits for $y^{0}$ are as in the previous section. Substituting the 
propagators (\ref{split}), using the Eqs. (\ref{OPERATOR1}), (\ref{OPERATOR2}), (\ref{OPERATOR3}), (\ref{OPERATOR4}), (\ref{OPERATOR5}), (\ref{OPERATOR6}), (\ref{OPERATOR7}), take into account only the contributions up to second order in
$v^{\mu}$, and evaluating the integrals: $d^{2}{\bf y}_{\perp}$, $d^{3}{\bf x}$, $dy^{3}$, $dp^{3}$, $dx^{0}$,
$dp^{0}$ and $dy^{0}$, we obtain
\begin{eqnarray}
\label{energy31} E_ {WT-LV}^{II}&=&2k^{6}M\mu\Bigl[\left({\bf{v}}_{\perp}\cdot{\bf{\nabla}}_{{\bf
a}_{\perp}}\right)^{2} \int\frac{d^{2}{\bf p}_{\perp}}{(2\pi)^{2}}\frac{e^{i{\bf p}_{\perp}\cdot{\bf
a}_{\perp}}}{{\bf p}^{2}_{\perp}} +\left[{\bf v}^{2}_{\perp}+2(v^{3})^{2}\right]\int\frac{d^{2}{\bf
p}_{\perp}}{(2\pi)^{2}}e^{i{\bf p}_{\perp}\cdot{\bf a}_{\perp}}\Bigr] \ ,
\end{eqnarray}
where $v^{3}$ is the projection of the background vector ${\bf{v}}$
along the string, and defined ${\bf p}_{\perp}=(p^{1},p^{2},0)$, ${\bf v}_{\perp}=(v^{1},v^{2},0)$, the distance between the particle and the cosmic string ${\bf
a}_{\perp}=(s^{1}-A^{1},s^{2}-A^{2},0)$, and the differential operator
\begin{eqnarray}
\label{deffo2} {\bf{\nabla}}_{{\bf a}_{\perp}}=\left(\frac{\partial}{\partial {a^{1}}},\frac{\partial}{\partial
{a^{2}}}, 0\right) \ .
\end{eqnarray}

Provided that ${{\bf a}_{\perp}}$ is non-zero, the last term inside the brackets in (\ref{energy31}) vanishes.
The remaining integral is divergent, in order to circumvent this problem we proceed as in references
\cite{LV8,fab1,fab2}, introducing a mass regulator parameter, as follows:
\begin{eqnarray}
\label{energy41} E_{WT-LV}^{II}&=&2k^{6}M\mu\left({\bf{v}}_{\perp}\cdot{\bf{\nabla}}_{{\bf a}_{\perp}}\right)^{2}
\lim_{m\rightarrow 0}\int\frac{d^{2}{\bf p}_{\perp}}{(2\pi)^{2}}\frac{e^{i{\bf p}_{\perp}\cdot{\bf a}_{\perp}}}
{{\bf p}^{2}_{\perp}+m^{2}} \ .
\end{eqnarray}

Using the fact that \cite{fab1}
\begin{eqnarray}
\label{int4EM} \int\frac{d^{2}{\bf q}_{\perp}}{(2\pi)^{2}}\frac{e^{i{\bf p}_{\perp}\cdot{\bf a}_{\perp}}}{{\bf
p}_{\perp}^2+m^{2}}=\frac{1}{2\pi} K_{0}\left(m{a}_{\perp}\right) \ ,
\end{eqnarray}
and acting with the differential operator (\ref{deffo2}), we arive at
\begin{eqnarray}
\label{energy51} E_{WT-LV}^{II}&=&\frac{k^{6}M\mu}{\pi}\lim_{m\rightarrow 0}\left[ \frac{\left({{\bf
v}_{\perp}}\cdot{{\bf a}_{\perp}}\right)^{2}}{{a}^{2}_{\perp}}m^{2}K_{2} \left(m{a}_{\perp}\right)-\frac{{{\bf
v}^{2}_{\perp}}}{{a}_{\perp}}mK_{1}\left(m{a}_{\perp}\right)\right] \ ,
\end{eqnarray}
where ${a}_{\perp}=\mid{\bf a}_{\perp}\mid$ and $K_{0}(m{{a}}_{\perp})$, $K_{1}(m{{a}}_{\perp})$,
$K_{2}(m{{a}}_{\perp})$ stand for the K-Bessel functions.

Using the fact that \cite{Arfken}
\begin{eqnarray}
\label{Kbessel} mK_{1}(m{{a}}_{\perp})\stackrel{m\rightarrow0}{\longrightarrow}\frac{1}{{{a}}_{\perp}} \ , \
m^{2}K_{2}(m{{a}}_{\perp})\stackrel{m\rightarrow0}{\longrightarrow}\frac{2}{{{a}}^{2}_{\perp}} \ ,
\end{eqnarray}
we obtain
\begin{eqnarray}
\label{energy61} E_{WT-LV}^{II}=\frac{512\pi^{2}G^{3}M\mu}{{a}_{\perp}^{2}}\left[ 2\frac{\left({{\bf
v}_{\perp}}\cdot{{\bf a}_{\perp}}\right)^{2}}{{a}^{2}_{\perp}}-{{\bf v}^{2}_{\perp}}\right]\ .
\end{eqnarray}

This interaction energy is an effect due solely to the Lorentz violating background up to lowest order of the
background vector, having no counterpart in standard WTDIFF theory. Clearly, if the background four-vector
$v^{\mu}$ is zero, there is no interaction energy. The same happens for the case where ${\bf{v}}_{\perp}=0$

The interaction force can be obtained from Eq. (\ref{energy61}) as follows,
\begin{eqnarray}
\label{Force62}
{\bf F}_{WT-LV}^{II}&=&-{\bf {\nabla}}_{{\bf a}_{\perp}}E_{WT-LV}^{II}\nonumber\\
&=&-\frac{1024\pi^{2}G^{3}M\mu}{{a}_{\perp}^{3}}\Biggl\{\left[ {{\bf v}^{2}_{\perp}}-4\frac{\left({{\bf
v}_{\perp}}\cdot{{\bf a}_{\perp}}\right)^{2}}{{a}^{2}_{\perp}} \right]{\hat{a}_{\perp}}+2\frac{\left({{\bf
v}_{\perp}} \cdot{{\bf a}_{\perp}}\right)}{{a}_{\perp}}{{\bf v}_{\perp}}\Biggr\} \ ,
\end{eqnarray}
where ${\hat{a}_{\perp}}$ is an unit vector pointing on the direction of vector ${{\bf a}}_{\perp}$.

As a final comment, we point out that from Eq. (\ref{energy61}), one can also obtain a torque on the cosmic
string by fixing the point particle.  Denoting by $\phi$ the angle between ${\bf v}_{\perp}$ and ${\bf
a}_{\perp}$, we obtain
\begin{eqnarray}
\label{TorqueII}
\tau_{WT-LV}^{II}&=&-\frac{\partial E_{WT-LV}^{II}}{\partial\phi}\nonumber\\
&=&\frac{1024\pi^{2}G^{3}M\mu}{{a}_{\perp}^{2}}{{\bf v}^{2}_{\perp}}\sin 2\left(\phi\right) \ .
\end{eqnarray}

We notice that the torque (\ref{TorqueII}) is an effect due solely to the Lorentz violating background. It does not appear in standard WTDIFF theory. If $\phi=0,\pi/2,\pi,2\pi$ or ${{\bf v}_{\perp}}=0$, the torque
vanishes. For $\phi=\pi/4$ the toque has a maximum intensity.

For LEH-LV theory we employ the same steps as above, we obtain for the interaction energy between the cosmic
string and the point-like particle the expression
\begin{eqnarray}
\label{energyEH66} 
E_{EH-LV}^{II}&=&\frac{k^{2}M\mu}{2t}\int\frac{d^{4}p}{\left(2\pi\right)^{4}}\int\int d^{4}x \ d^{4}y \ \delta^{3}\left({\bf x}-{\bf s}\right)\delta^{2}\left({\bf{y}}_{\perp}-{\bf A}\right)e^{-ip\cdot\left(x-y\right)}\Bigl[D_{00,00}^{(0)}\left(p\right)-D_{00,33}^{(0)}\left(p\right)\nonumber\\
&
&+D_{00,00}^{(LV)}\left(p\right)-D_{00,33}^{(LV)}\left(p\right)\Bigr]
=4E_{WT-LV}^{II} \ ,
\end{eqnarray}
showing that we have a numerically different result for both WTDIFF-LV and LEH-LV theories
although they can be related via a scaling of the Lorentz breaking term once again.

The next and last example is given by two parallel cosmic strings. We take a coordinate system where the first
string lies along the straight line located at ${\bf A}_{1}=(A_{1}^{1},A_{1}^{2},0)$, with linear mass density
$\mu_{1}$, and the second string lies along the line that crosses the $xy$ plane at ${\bf
A}_{2}=(A_{2}^{1},A_{2}^{2},0)$, with linear mass density $\mu_{2}$. The corresponding external source is given
by
\begin{eqnarray}
\label{corre3} \left[T^{\alpha\beta}({x})\right]^{III}= \mu_{1}\left(\eta^{\alpha 0}\eta^{\beta 0}-\eta^{\alpha
3}\eta^{\beta 3}\right)\delta^{2}\left({\bf{x}}_{\perp}-{{\bf{ A}}_{1}}\right)+\mu_{2}\left(\eta^{\alpha
0}\eta^{\beta 0}-\eta^{\alpha 3}\eta^{\beta 3}\right)\delta^{2}\left({\bf{x}}_{\perp}-{{\bf{A}}_{2}}\right) \ ,
\end{eqnarray}
\begin{eqnarray}
\label{corre33} \left[{\hat{T}}^{\alpha\beta}({x})\right]^{III}= \mu_{1}\left(\eta^{\alpha 0}\eta^{\beta
0}-\eta^{\alpha 3}\eta^{\beta 3}-\frac{\eta^{\alpha\beta}}{2}\right)\delta^{2}\left({\bf{x}}_{\perp}-{\bf{
A}}_{1}\right)+\mu_{2}\left(\eta^{\alpha 0}\eta^{\beta 0}-\eta^{\alpha 3}\eta^{\beta
3}-\frac{\eta^{\alpha\beta}}{2}\right)\delta^{2}\left({\bf{x}}_{\perp}-{\bf{ A}}_{2}\right)\ .
\end{eqnarray}

Substituting the sources (\ref{corre3}) and (\ref{corre33}) in (\ref{energy1WT3}),
 discarding the
self-interacting terms, using the Eqs. (\ref{split}), (\ref{OPERATOR1}), (\ref{OPERATOR2}), (\ref{OPERATOR3}), (\ref{OPERATOR4}), (\ref{OPERATOR5}), (\ref{OPERATOR6}), (\ref{OPERATOR7}),
 proceeding as in the previous
cases, and identifying the length of the cosmic string as $L=\int dx^{3}$, we can show that the interaction
energy between the two cosmic strings up to second order in $v^{\mu}$ is given by
\begin{eqnarray}
\label{energy41dd} E_{WT-LV}^{III}&=&4k^{6}\mu_{1}\mu_{2}L\left({\bf{v}}_{\perp}\cdot{\bf{\nabla}}_{{\bf
a}_{\perp}}\right)^{2} \lim_{m\rightarrow 0}\int\frac{d^{2}{\bf p}_{\perp}}{(2\pi)^{2}}\frac{e^{i{\bf
p}_{\perp}\cdot{\bf a}_{\perp}}}
{{\bf p}^{2}_{\perp}+m^{2}} \nonumber\\
&=&\frac{1024\pi^{2}G^{3}\mu_{1}\mu_{2}L}{{a}_{\perp}^{2}}\left[ 2\frac{\left({{\bf v}_{\perp}}\cdot{{\bf
a}_{\perp}}\right)^{2}}{{a}^{2}_{\perp}}-{{\bf v}^{2}_{\perp}}\right]\ ,
\end{eqnarray}
where we have identified the distance between the strings as ${\bf{a}}_{\perp}={\bf{A}}_{1}-{\bf{A}}_{2}$, and $a_{\perp}=\mid{\bf{a}}_{\perp}\mid$ .

It can be seen that the energy given above vanishes in the limit $v^{\mu}=0$, where we do not have
Lorentz-symmetry breaking. 

Similarly, the energy (\ref{energy41dd}) leads to an interaction force between two cosmic strings as well as to a 
torque on one string when we fix the other one.

For the LEH-LV theory, we obtain the expression
\begin{eqnarray}
\label{energyEH663} E_{EH-LV}^{III}=4E_{WT-LV}^{III} \ ,
\end{eqnarray}
where the numerical difference between the results for both theories was already expected.

\section{Conclusions and perspectives}
\label{conclusoes}

In this paper we have investigated the interactions between stationary sources for the WTDIFF and LEH theories
in the presence of the Lorentz violating and CPT breaking topological Chern-Simons term in $3 + 1$ dimensions.  All the results
have been obtained perturbatively up to second order in $v^{\mu}$. We have shown the emergence of an spontaneous
torque on a classical rigid halter. We have also investigated interactions with one or two cosmic strings. We have shown
that a cosmic string has a non trivial interaction with a point-like particle as well as with another cosmic
string. All those phenomena would not appear in the absence of the Lorentz violating term and
 have not been explored before in the literature.

We have also shown that there is a numerical difference
in the interactions between stationary external sources regarding
WTDIFF-LV and LEH-LV theories. At leading order in perturbation theory the differences
can be extinguished after a convenient scaling of the Lorentz violating source term which
makes their experimental detection impossible in the cases analyzed here. It is not clear however
if such scaling will keep working beyond the leading order in perturbation theory. This is
under investigation as well as  the investigation of more general non minimal 
Lorentz violating term, see \cite{G15}.

\section{Appendix}

Here we give a summary of technical details about the propagator found in \cite{G3} which include a class of
Lorentz violating terms. We use a slightly different notation. The Lagrangian densities for symmetric rank-2
fields can be written as ${\cal L} = h_{\alpha\beta}G^{\alpha\beta,\mu\nu}h_{\mu\nu}$. Let us assume that the
differential operator $G^{\alpha\beta,\mu\nu}$, suppressing indices, can be decomposed as

\be G = A \,  P_{SS}^{(2)}  + B \,  P_{SS}^{(1)} + A_{SS}  P_{SS}^{(0)} + A_{WW}  P_{WW}^{(0)} + A_{SW}(
P_{WS}^{(0)} + P_{SW}^{(0)}) + a\,  S_{\tau} + b \, S_{\sigma} + c\, S_{\tau}^2 + d\, S_{\sigma}^2 \quad .
\label{G} \ee

\no  where the spin-s operators $P_{IJ}^{(s)}$ acting on symmetric rank-2 tensors in $D=4$ are given by

\be \left( P_{SS}^{(2)} \right)^{\lambda\mu}_{\s\s\alpha\beta} = \frac 12 \left(
\theta_{\s\alpha}^{\lambda}\theta^{\mu}_{\s\beta} + \theta_{\s\alpha}^{\mu}\theta^{\lambda}_{\s\beta} \right) -
\frac{\theta^{\lambda\mu} \theta_{\alpha\beta}}{3} \quad , \label{ps2} \ee

\be \left( P_{SS}^{(1)} \right)^{\lambda\mu}_{\s\s\alpha\beta} = \frac 12 \left(
\theta_{\s\alpha}^{\lambda}\,\omega^{\mu}_{\s\beta} + \theta_{\s\alpha}^{\mu}\,\omega^{\lambda}_{\s\beta} +
\theta_{\s\beta}^{\lambda}\,\omega^{\mu}_{\s\alpha} + \theta_{\s\beta}^{\mu}\,\omega^{\lambda}_{\s\alpha}
\right) \quad , \label{ps1} \ee

\be \left( P_{SS}^{(0)} \right)^{\lambda\mu}_{\s\s\alpha\beta} = \frac 1{3} \,
\theta^{\lambda\mu}\theta_{\alpha\beta} \quad , \quad \left( P_{WW}^{(0)} \right)^{\lambda\mu}_{\s\s\alpha\beta}
= \omega^{\lambda\mu}\omega_{\alpha\beta} \quad , \label{psspww} \ee

\be \left( P_{SW}^{(0)} \right)^{\lambda\mu}_{\s\s\alpha\beta} = \frac 1{\sqrt{3}}\,
\theta^{\lambda\mu}\omega_{\alpha\beta} \quad , \quad  \left( P_{WS}^{(0)}
\right)^{\lambda\mu}_{\s\s\alpha\beta} = \frac 1{\sqrt{3}}\, \omega^{\lambda\mu}\theta_{\alpha\beta} \quad ,
\label{pswpws} \ee

\be  \omega_{\mu\nu} = \frac{\p_{\mu}\p_{\nu}}{\Box} \quad , \quad \theta_{\mu\nu} = \eta_{\mu\nu} -
\frac{\p_{\mu}\p_{\nu}}{\Box}\quad , \label{pvectors} \ee

\no while in the Lorentz-violating sector we have the spin-2 operators

\begin{eqnarray}
\label{OPERATOR1}
\left(S_{\tau}\right)_{\alpha\beta,\gamma\delta}&=&\frac{1}{2}\left(\tau_{\alpha\gamma}
S_{\beta\delta}+\tau_{\alpha\delta}S_{\beta\gamma}+\tau_{\beta\gamma}
S_{\alpha\delta}+\tau_{\beta\delta}S_{\alpha\gamma}\right) \ , \\
\label{OPERATOR2}
\left(S_{\tau}^{2}\right)_{\alpha\beta,\gamma\delta}&=&f\left(p\right)
\left[\frac{1}{2}\left(\theta_{\alpha\gamma}\tau_{\beta\delta}+\theta_{\alpha\delta}\tau_{\beta\gamma}+
\theta_{\beta\gamma}\tau_{\alpha\delta}+\theta_{\beta\delta}\tau_{\alpha\gamma}\right)-
\left(\tau_{\alpha\gamma}\tau_{\beta\delta}+\tau_{\alpha\delta}\tau_{\beta\gamma}\right)\right] \ , \\
\label{OPERATOR3}
\left(S_{\sigma}\right)_{\alpha\beta,\gamma\delta}&=&\frac{1}{2}\left(\sigma_{\alpha\gamma}
S_{\beta\delta}+\sigma_{\alpha\delta}S_{\beta\gamma}+\sigma_{\beta\gamma}
S_{\alpha\delta}+\sigma_{\beta\delta}S_{\alpha\gamma}\right) \ , \\
\label{OPERATOR4}
\left(S_{\sigma}^{2}\right)_{\alpha\beta,\gamma\delta}&=&f\left(p\right)
\left[\sigma_{\alpha\gamma}\sigma_{\beta\delta}+\sigma_{\alpha\delta}\sigma_{\beta\gamma}-
\left(S_{\alpha\gamma}S_{\beta\delta}+S_{\alpha\delta}S_{\beta\gamma}\right)\right]
\ ,
\end{eqnarray}

\no with

\begin{eqnarray}
 \label{OPERATOR5}
\tau_{\alpha\beta}&=&\frac{1}{f\left(p\right)}\left[\left(v\cdot p\right)\left(v_{\alpha}p_{\beta}+v_{\beta}p_{\alpha}\right)-p^{2}v_{\alpha}v_{\beta}-\frac{\left(v\cdot p\right)^{2}p_{\alpha}p_{\beta}}{p^{2}}\right] \ , \\
\label{OPERATOR6}
 \sigma_{\alpha\beta}&=&\theta_{\alpha\beta}-\tau_{\alpha\beta} \ , \ S_{\alpha\beta}=k^{2}\varepsilon_{\alpha\beta\rho\phi}v^{\rho}p^{\phi} \ , \\
\label{OPERATOR7} f\left(p\right)&=&k^{4}\left[\left(v\cdot p\right)^{2}-v^{2}p^{2}\right] \quad ; \quad p_{\mu}
= - \imath \, \p_{\mu} \quad . \label{efe}
\end{eqnarray}

\no From the algebra

\bea \left(P^{(s)}\right)_{IJ} \left(P^{(r)}\right)^{JK} &=& \delta^{rs} \left(P^{(s)}\right)_I^{\, \, K} \quad
;\quad S_{\sigma} P^{(s)}_{IJ}= \delta^{s2}  S_{\sigma} \, ; \, S_{\tau} P^{(s)}_{IJ}= \delta^{s2}  S_{\tau}
\label{algebra1} \\ S_{\sigma} \cdot S_{\tau} &=& 0 \, ; \, S_{\sigma}^3 = 4 \, f \, S_{\sigma} \quad ; \quad
S_{\tau}^3 =  f \, S_{\tau}  \quad . \label{algebra2} \eea

\no The reader can check that the inverse of the operator (\ref{G}) is \cite{G3} :

\bea G^{-1} &=&  \frac{ P_{SS}^{(2)}}A  +  \frac{P_{SS}^{(1)}}B + \frac{A_{WW}  P_{SS}^{(0)} + A_{SS}
P_{WW}^{(0)} - A_{SW}( P_{WS}^{(0)} + P_{SW}^{(0)})}{K^{(0)}} \nn\\ && + \frac a{D_1} S_{\tau} + \frac b{D_2}
S_{\sigma} + \frac{ c(A+c \, f)- a^2}{A\, D_1} S_{\tau}^2 + \frac{ d(A+ 4\, d \, f)- b^2}{A\, D_2} S_{\sigma}^2
\quad . \label{Gmenos} \eea

\no where $f$ is given in (\ref{efe}) and

\be D_1 = a^2 \, f - (A + c \, f)^2 \, ; \, D_2 = 4\, b^2 \, f - (A + 4 \, d\, f)^2 \, ; \, K^{(0)} = A_{WW}\, A_{SS}
- A_{WS}^2 \quad . \label{d1d2} \ee

\begin{acknowledgments}
L. H. C. Borges thanks the S\~ao Paulo Research Foundation (FAPESP), grant 2016/11137-5, for financial
support. D. Dalmazi thanks CNPq (Brazilian agency), grant 306380/2017-0 for financial support.
\end{acknowledgments}



\end{document}